\documentclass[aps,prl,reprint,superscriptaddress,floatfix]{revtex4-1}
\usepackage{amsmath,latexsym}
\usepackage{color}
\usepackage[colorlinks=true,allcolors=blue]{hyperref}
\usepackage{graphicx}

\makeindex

\begin{document}

\title{Temperature dependent relaxation of dipole-exchange magnons \\ in yttrium iron garnet films}

\author{Laura Mihalceanu}
\email{mihalcea@rhrk.uni-kl.de}
\affiliation{Fachbereich Physik and Landesforschungszentrum OPTIMAS, Technische Universit\"at Kaiserslautern, 67663 Kaiserslautern, Germany \looseness=-1}

\author{Vitaliy I.~Vasyuchka}
\affiliation{Fachbereich Physik and Landesforschungszentrum OPTIMAS, Technische Universit\"at Kaiserslautern, 67663 Kaiserslautern, Germany \looseness=-1}

\author{Dmytro A.~Bozhko}
\affiliation{Fachbereich Physik and Landesforschungszentrum OPTIMAS, Technische Universit\"at Kaiserslautern, 67663 Kaiserslautern, Germany \looseness=-1}

\author{Thomas~Langner}
\affiliation{Fachbereich Physik and Landesforschungszentrum OPTIMAS, Technische Universit\"at Kaiserslautern, 67663 Kaiserslautern, Germany \looseness=-1}

\author{Alexey~Yu.~Nechiporuk}
\affiliation{Faculty of Radiophysics, Electronics and Computer Systems, Taras Shevchenko National University of Kyiv, 01601 Kyiv, Ukraine \looseness=-1}

\author{Vladyslav F. Romanyuk}
\affiliation{Faculty of Radiophysics, Electronics and Computer Systems, Taras Shevchenko National University of Kyiv, 01601 Kyiv, Ukraine \looseness=-1}

\author{Burkard~Hillebrands}
\affiliation{Fachbereich Physik and Landesforschungszentrum OPTIMAS, Technische Universit\"at Kaiserslautern, 67663 Kaiserslautern, Germany \looseness=-1}

\author{Alexander A.~Serga}
\affiliation{Fachbereich Physik and Landesforschungszentrum OPTIMAS, Technische Universit\"at Kaiserslautern, 67663 Kaiserslautern, Germany \looseness=-1}

\date{\today}

\begin{abstract}

Low energy consumption enabled by charge-free information transport, which is free from ohmic heating, and the ability to process phase-encoded data by nanometer-sized interference devices at GHz and THz frequencies are just a few benefits of spin-wave-based technologies. Moreover, when approaching cryogenic temperatures, quantum phenomena in spin-wave systems pave the path towards quantum information processing. 
In view of these applications, the lifetime of magnons---spin-wave quanta---is of high relevance for the fields of magnonics, magnon spintronics and quantum computing.
Here, the relaxation behavior of parametrically excited magnons having wavenumbers from zero up to $6\cdot 10^5\,\mathrm{rad\,cm}^{-1}$ was experimentally investigated in the temperature range from 20\,K to 340\,K in single crystal yttrium iron garnet (YIG) films epitaxially grown on gallium gadolinium garnet (GGG) substrates as well as in a bulk YIG crystal---the magnonic materials featuring the lowest magnetic damping known so far. 
As opposed to the bulk YIG crystal in YIG films we have found a significant increase in the magnon relaxation rate below 150\,K---up to 10.5 times the reference value at 340\,K---in the entire range of probed wavenumbers. 
This increase is associated with rare-earth impurities contaminating the YIG samples with a slight contribution caused by coupling of spin waves to the spin system of the paramagnetic GGG substrate at the lowest temperatures. \looseness=-1

\end{abstract}

\pacs{}

\keywords{}

\maketitle {}






\newpage
\thispagestyle{empty}

The fields of spintronics and magnonics promote the realization of faster data processing technologies with lower energy dissipation by complementing or even replacing electron charge-based technologies with spin degree of freedom based devices \cite{Sato, Kruglyak2010, Chumak2015}. Simultaneously, novel fascinating magnetic phenomena---such as, e.g., room-temperature Bose-Einstein magnon condensates \cite{Demokritov2006, Serga2014, Safranski2017}, magnon vortices \cite{Nowik-Boltyk2012} and supercurrents \cite{Bozhko2016, Nakata2014, Skarsvag2015, Flebus2016}---open a whole new range of research areas \cite{Bauer2012, Chumak2015} both for basic and applied spin physics. For these purposes many novel materials have been designed and investigated \cite{Harris2012, Felser2007, Hirohata2015} whereupon one of the most outstanding ones so far is the insulating ferrimagnet yttrium iron garnet ($\mathrm{Y_{3}Fe_{5}O_{12}}$, YIG). \looseness=-1

Since its discovery in 1956, YIG has served as a prime example material for its microwave, optical, acoustic, and magneto-optical properties \cite{Cherepanov1993} in a wide range of experiments and applications. Nowadays, single crystal YIG films epitaxially grown on gadolinium gallium garnet ($\mathrm{Gd_{3}Ga_{5}O_{12}}$, GGG) substrates \cite{Dubs2017, Sun2012, Chang2014} dominate in theoretical and experimental studies \cite{Serga2014, Safranski2017, Nowik-Boltyk2012, Bozhko2016, Hamadeh2014, Yu2014, Cornelissen2015, Schreier2016, Barker2016}.  Their pertinence ranges from building of devices like microwave YIG oscillators, filters, delay lines, phase shifters, etc. \cite{Adam1988} up to the latest high-profile research as in magnonics \cite{Serga2010, Demokritov}, spintronics \cite{Sato, Hellman2017} and quantum computing \cite{Andrich2017, Tabuchi2016}. Consequently it has become clear that a deep understanding of the magnetic damping properties, determining the magnon lifetimes, is of crucial importance throughout these fields. 
Given its high Curie temperature at 560\,K, YIG is applicable at ambient temperatures, where its exceptional low Gilbert damping parameter of down to $10^{-5}$ \cite{Klingler2017} enables a long spin precession lifetime. Furthermore, in quantum computing YIG is used at very low temperatures for the coupling of single magnons to superconductive qubits in microwave cavities for the storage of information \cite{Tabuchi2015, Tabuchi2014, Huebl2013, Kosen2017, Morris2017}.
All of this in connection with arising demands on miniaturization of magnonic devices motivates our studies of the damping behavior in YIG films towards cryogenic temperatures in a wide range of spin-wave wavelengths.  \looseness=-1

Up to now the temperature dependence of magnetic damping in YIG has been examined only for long-wavelength dipolar magnons with wavenumbers $q \rightarrow 0$ \cite{Spencer1961, Vittoria1985}. There exist numerous ways to measure the relaxation behavior of a precessing magnetic moment in different ranges of magnon wavenumbers. Among the most established are the technique of ferromagnetic resonance (FMR) \cite{Kalarickal2006}, the measurement of thresholds of parametric excitation of magnetic oscillations and waves \cite{Kasuya1961}, the determination of spin-wave relaxation time from direct observation of magnetization decay by means of time-resolved Brillouin light scattering spectroscopy \cite{Serga2014,Jungfleisch2011,Sebastian2012} and the magneto optical Kerr effect \cite{Razdolski2017}, the magnetic-resonance force microscopy \cite{Klein2003}, and echo-methods \cite{Kaplan1966, Danilov1983,Melkov2004}. However, only parametric excitation techniques allow for the effective excitation and probing of dipolar-exchange magnons with wavenumbers up to $q \leq 10^6\,\mathrm{rad\,cm}^{-1}$ \cite{Gurevich1975, Langner2017}. For example, the parametric pumping process of the first order, as described by Suhl \cite{Suhl1957} and Schl\"{o}mann \textit{et al.} \cite{Schloemann1960}, resembles a form of spin-wave excitation when either a magnon of an externally driven magnetization precession or a photon of a pumping microwave magnetic field with wavenumbers $q_{\mathrm{p}}\approx 0$ splits into two magnons with opposite wavevectors $\mathbf{q}$ and $-\mathbf{q}$ at half of the pumping frequency $\omega_{\mathrm{p}}/2$. Thus, a rather spatially uniform microwave magnetic field can generate short-wavelength magnons, whose wavenumbers are determined by the applied bias magnetic field $H$ and are bounded above only by the chosen pumping frequency $\omega_{\mathrm{p}}$.  \looseness=-1

\begin{figure}[t]
\includegraphics[width = 1.0\columnwidth]{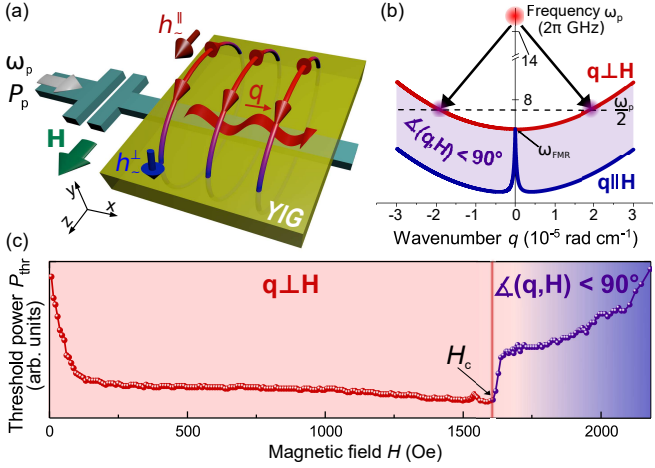}
\caption{\label{fig1} (a) Sketch of the experimental setup. The spin system of a YIG sample, which is placed on top of a microstrip resonator, is driven by a microwave Oersted field with components oriented perpendicular ($\mathbf{h}^{\perp}_{\sim}$) and parallel ($\mathbf{h}^{\parallel}_{\sim}$) to the bias magnetic field $\mathbf{H}$. (b) Schematic illustration of the parametric pumping process in an in-plane magnetized YIG film. The transversal (red curve) and longitudinal (blue curve) lowest magnon branches are calculated for $H=1600\,\mathrm{Oe}$. The purple area contains the magnon branches with wavevectors lying in the film plane in the angle range between 0 and $90^\circ$ relative to the field $\mathbf{H}$. Two arrows show the splitting of a microwave photon in two  magnons at half of the pumping frequency $\omega_{\mathrm{p}}/2$. For the given bias magnetic field the magnons are excited on the transversal dispersion branch. (c) Dependence of the threshold power $P_{\mathrm{thr}}$ of parametric instability on the bias magnetic field $H$ measured in a 53\,$\mathrm{\mu m}$-thick YIG film at 60\,K. The minima of the threshold curve at $H=H_{\mathrm{c}}$ corresponds to the excitation of magnons with wavenumbers $q \rightarrow 0$ near the frequency of the ferromagnetic resonance. At $H<H_{\mathrm{c}}$ dipolar-exchange magnons corresponding to the transversal dispersion branch are directly excited by the parallel component $\mathbf{h}^{\parallel}_{\sim}$ of the pumping Oersted field. For $H>H_{\mathrm{c}}$ the magnons from the purple spectral area (panel (b)) are excited by the precessing magnetization driven by the perpendicular component $\mathbf{h}^{\perp}_{\sim}$ of the pumping Oersted field. \looseness=-1
}
\end{figure}

In our experiments, we investigated parametrically excited magnons in in-plane magnetized YIG films of thicknesses of 5.6~$\mathrm{\mu m}$, 6.7~$\mathrm{\mu m}$ and 53~$\mathrm{\mu m}$, which were epitaxially grown in the (111) crystallographic plane on GGG substrates of $500\,\mathrm{\mu m}$ thickness. In addition, the GGG substrate was mechanically polished away from the originally 53\,$\mathrm{\mu m}$-thick sample down to a 30\,$\mathrm{\mu m}$-thick YIG film. This sample was used to reveal a possible contribution of the interaction between the ferrimagnetic YIG and paramagnetic GGG spin systems to the magnon damping. The YIG samples with lateral sizes of $1\times5\,\mathrm{mm^2}$ prepared by chemical etching on the $5\times6\,\mathrm{mm^2}$ large GGG substrates were magnetized along their long axis to avoid undesirable influence of static demagnetizing on the value of the internal magnetic field.

The experimental realization is provided by the microwave setup shown in Fig.\,\ref{fig1}(a). The setup is attached on a highly heat-conducting $\mathrm{AlN}$ substrate at the bottom and is allocated inside a closed cycle refrigerator system.
A microwave pumping pulse of 10\,$\mu$s duration at a frequency $\omega_{\mathrm{p}}$ of $2 \pi \cdot 14$\,{GHz} with a 10\,ms repetition rate and a maximal pumping power $P_{\mathrm{p}}$ of 12\,{W} feeds a 50\,$\mu$m-wide microstrip resonator capacitively coupled to a microwave transmission line. The microwave Oersted field ${\bf h_{\mathrm{p}}}$ induced by the resonator drives the magnetization of a YIG-film sample placed on top of the microstrip. 
Subsequently the signal reflected by the resonator is forwarded to an oscilloscope.

When the threshold field condition $h_{\mathrm{p}} = h_{\mathrm{thr}}$ is fulfilled, the action of the microwave Oersted field compensates the spin-wave damping and gives rise to the parametric instability process, where a selected magnon mode, which has the lowest damping and the strongest coupling to the pumping, grows exponentially in time. The arising mode increasingly absorbs the microwave energy accumulated in the pump resonator. This process detunes the resonator and, thus, changes the level of the reflected signal passed to the oscilloscope. As a result, a kink appearing at the end of the reflected pump pulse indicates the threshold microwave power $P_\mathrm{p}=P_\mathrm{thr}$ required for the parametric excitation process \cite{Neumann2009}. $P_\mathrm{thr}$ can be determined for magnon modes over the wide $q$-spectral range by changing the magnetic field ${H}$, which leads to a vertical shift of the dispersion curve (Fig.\,\ref{fig1}\,(b)) along frequency axis and results in a characteristic threshold curve shown in Fig.\,\ref{fig1}\,(c).  

\begin{figure*}[t]
	\includegraphics[width =1.8\columnwidth]{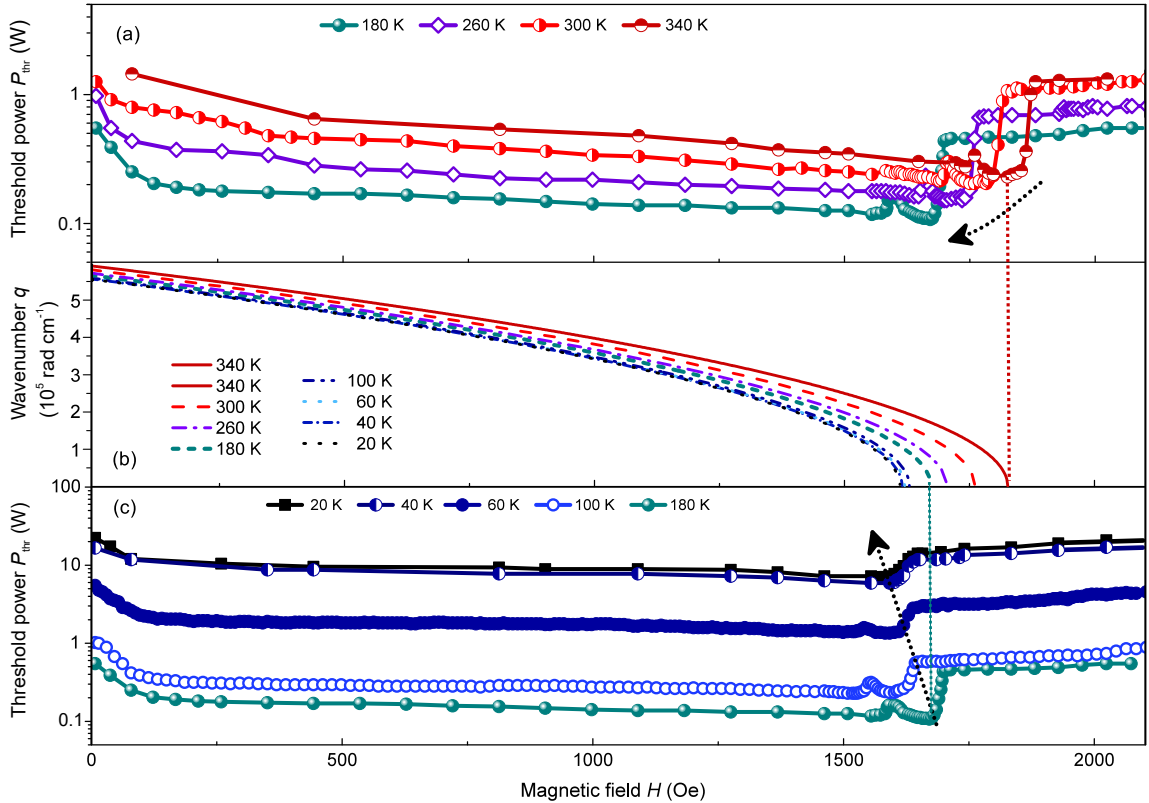}
	\caption{\label{fig2}  Threshold curves $P_{\mathrm{thr}}(H)$ at different temperatures in the range $340-180$\,K (a) and $180-20$\,K (c). (b) Wavenumber in the wide temperature range. All present data is recorded and calculated for a 53\,$\mathrm{\mu m}$-thick YIG film grown on top of a GGG substrate. \looseness=-1}
\end{figure*}

In order to understand the shape of this curve one needs to consider that the overall threshold power $P_\mathrm{thr}$ is determined by instabilities of magnons excited by the components of the microwave Oersted field oriented both perpendicular $\mathbf{h}^\perp_\sim$ (blue arrow in Fig.~\ref{fig1}\,(a)) and parallel $\mathbf{h}^\parallel_\sim$ (red arrow in Fig.~\ref{fig1}\,(a)) to the bias magnetic field $\mathbf{H}$ \cite{Neumann2009}. At the critical field $H=H_\mathrm{c}$ spin waves with $q \rightarrow 0$ are excited near the frequency of the ferromagnetic resonance: $\omega_{\mathrm{p}}/2 \approx \omega_\mathrm{FMR}$. In Fig.\,\ref{fig1}\,(c) this situation corresponds to the minima of the threshold curve $P_\mathrm{thr} (H)$. The threshold power at $H \leq H_\mathrm{c}$ is dominated by direct parametric interaction of the parallel field component $\mathbf{h}^\parallel_\sim$ with the lowest thickness mode corresponding to the transversal magnon dispersion branch (red curve in Fig.~\ref{fig1}\,(b)) \cite{Serga2012}. As this mode is characterized by the largest precession ellipticity, the longitudinal component $\mathbf{m}_z$ of the precessing magnetic moment strongly oscillates along the direction of the magnetic field $\mathbf{H}$ with frequency $\omega_{\mathrm{p}}$ and thus effectively couples with the parallel component $\mathbf{h}^\parallel_\sim$ of the pumping field. 
With decreasing external magnetic field, the threshold power slowly increases due to an increase in wavenumbers of the excited magnons and a related decrease in the precession ellipticity \cite{Serga2012}.
The strong increase in $P_\mathrm{thr}$ at the magnetic field $H$ below 100\,Oe is caused by transition of the homogeneously magnetized YIG film to a multi-domain state.

Above $H_\mathrm{c}$ no magnons with wavevectors $\mathbf{q} \perp \mathbf{H}$ exist at $\omega_{\mathrm{p}} / 2$ and the parametric pumping excites magnons propagating at angles $\theta_q < 90^\circ$ relative to the field $\mathbf{H}$. These magnons escape the narrow pumping area above the microstrip resonator (Fig.~\ref{fig1}\,(a)) and the related energy leakage results in the sharp jump up in the threshold power just above $H_{\mathrm{c}}$. This confinement effect together with general reduction in the precession ellipticity caused by the decrease of $\theta_q$ leads to a further transition from the parallel to the perpendicular pumping regimes for $H > H_\mathrm{c}$ \cite{Neumann2009}. Finally, the threshold power $P_\mathrm{thr} \rightarrow \infty$ when the bottom of the magnon spectrum is shifted above $\omega_{\mathrm{p}}/2$.

For determining the magnon relaxation behavior the pumping regime $H \leq H_\mathrm{c}$ is of main interest in this report as the wavenumbers of the parametrically excited magnons can be unambiguously calculated in this case. Henceforth we approximate $h_\mathrm{p} \simeq h^\parallel_\sim$.

Figure~\ref{fig2} presents the dependencies $P_\mathrm{thr} (H)$ recorded for a number of temperatures in the range from 340\,K to 180\,K (Fig.\,\ref{fig2}\,(a)) and from 180\,K to 20\,K (Fig.\,\ref{fig2}\,(c)) for the 53\,$\mathrm{\mu m}$-thick film. The dotted arrows indicate the shift of both the critical threshold power $P_\mathrm{thr}(H_\mathrm{c})$ and $H_\mathrm{c}$ with temperature. One can see that Fig.~\ref{fig2}\,(a) shows a \textit{decrease} in the threshold power with decreasing temperature from 340\,K to 180\,K. On the contrary, Fig.\,\ref{fig2}\,(c) reveals a strong \textit{increase} in the threshold power with further temperature decrease from 180\,K to 20\,K.
At the same time, the experimentally determined critical field $H_{\mathrm{c}}$ monotonically decreases towards lower temperatures along the whole temperature range. 

This decrease of $H_{\mathrm{c}}$ relates to an upward frequency shift of the magnon spectrum caused by a temperature dependent increase in the saturation magnetization $4 \pi M_{\mathrm{s}}$ as well as by changes of the cubic $H^\mathrm{c}_\mathrm{a}$ and uniaxial $H^\mathrm{u}_\mathrm{a}$ anisotropy fields of the YIG film \cite{Bobkov2003, Gurevich}. The field dependence of the wavevector spectral range for the perpendicular spin-wave branch can be calculated using Eq.\,7.9 from Ref.~\cite{Gurevich}:
\begin{equation}
\omega=\gamma\sqrt{(H+D q^{2})(H+D q^{2}+4 \pi M_\mathrm{s} - H^\mathrm{c}_\mathrm{a} - H^\mathrm{u}_\mathrm{a})},
\label{k_wave}
\end{equation} 
where $\omega=\omega_{\mathrm{p}} / 2$, $\gamma = 1.76\cdot10^{7}\,\mathrm{Oe^{-1}\, s^{-1}}$ the gyromagnetic ratio, and the nonuniform exchange constant $D=5.2\cdot10^{-9}$\,Oe\,$\mathrm{cm^{2}}$  are considered to be not varying with temperature \cite{LeCraw1961}. The difference $4 \pi M_{\mathrm{s}} - H^\mathrm{c}_\mathrm{a} - H^\mathrm{u}_\mathrm{a}$ is defined from the measured values of $H_\mathrm{c} (T,q=0)$. An expected demagnetizing effect caused by a stray magnetic field induced at low temperatures in YIG films by the paramagnetic GGG substrate can be neglected in our case of laterally extended samples \cite{Danilov1989}.
%

The calculated dependencies of the magnon wavenumber $q=q(H)$ for different temperatures are shown in Fig.\,\ref{fig2}\,(b). 
The vertical dashed lines in Fig.\,{\ref{fig2}} correlate the threshold curves with the corresponding wavenumber at $H_{\mathrm{c}}$. As is shown, in our experiment spin waves are probed by parametric pumping in the wavenumber range from zero to $6\cdot 10^5\,\mathrm{rad\,cm}^{-1}$. 

%

The variation of the saturation magnetization directly affects the coupling between the microwave pumping field $ {\vec{h}{^{\parallel}_{\sim}}} $ and the longitudinal component $\mathbf{m}_z$ of the precessing magnetic moment $\bf{M}$.
As a result, the threshold field $ h_{\mathrm{thr}}$ is influenced by two temperature dependent physical quantities: the spin-wave relaxation rate and the parametric coupling strength.
These influences can be estimated using the relation for the threshold field \cite{Gurevich}:
\begin{equation} 
h_{\mathrm{thr}} = \mathrm{min} \left\{ \frac {\omega_{\mathrm{p}} \Delta H_{q}} {\omega_{\mathrm{M}} \sin^2\theta_{q}} \right\},
\label{h_thr}
\end{equation}
where $\omega_{\mathrm{M}} = \gamma 4 \pi M_{\mathrm{s}}$, and $\theta_q$ is the angle between the magnon wavevector $\bf{q}$ and the magnetization direction. For the parametric excitation near and above the FMR frequency ($H\le H_\mathrm{c}$), we can approximate $\theta_{q} \approx 90^\circ$. 
$\Delta H_{q}$ is the width of a linear resonance curve of the parametrically excited magnon mode with the wavenumber $q$. It is defined as $\Delta H_{q}=1 / (\gamma T_{q})=\Gamma_{q} / \gamma$, where $T_{q}$ and $\Gamma_{q}$ are the magnon lifetime and the spin-wave relaxation rate. 

\begin{figure*}[t]
	\includegraphics[width =1.8\columnwidth]{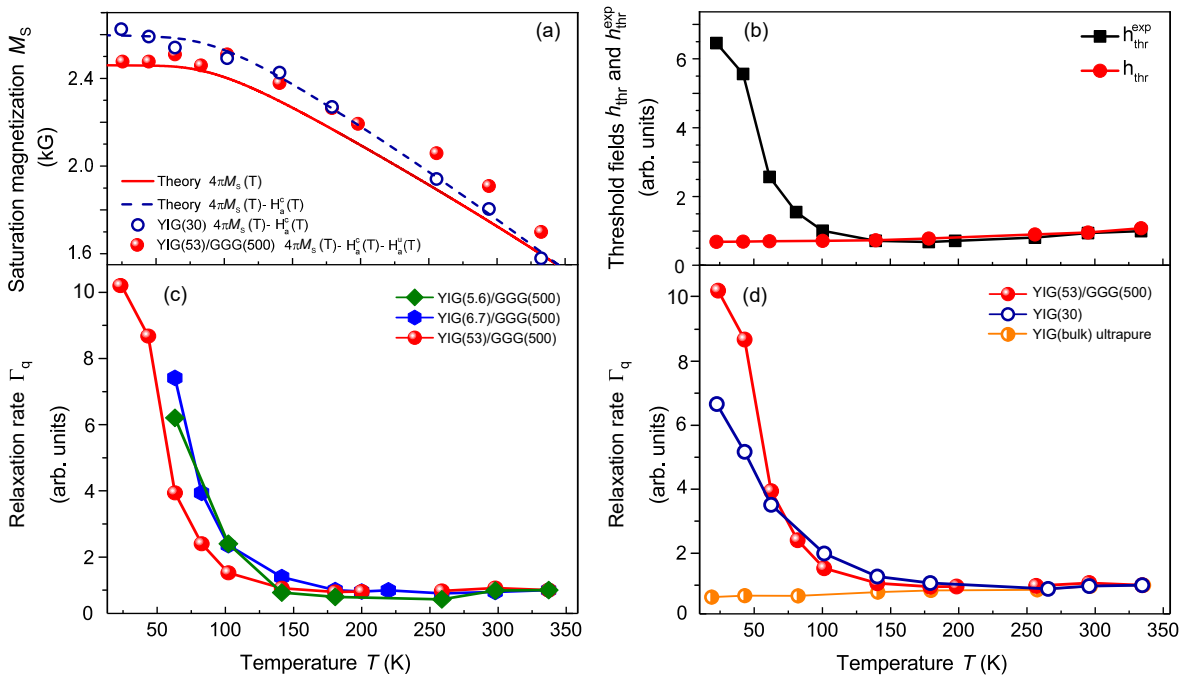}
	\caption{\label{fig3} (a) Saturation magnetization plotted as a function of temperature compared to theoretical calculations. (b) Temperature dependence of the threshold pumping field for magnons parametrically excited near the FMR frequency at $H=H_\mathrm{c}$. Squares -- the $h_{\mathrm{thr}}^\mathrm{exp}$ values are determined using the measured threshold powers $P_\mathrm{thr}$. Circles -- the $h_{\mathrm{thr}}$ values are calculated using Eq.~\ref{h_thr} for the experimentally determined values of $4\pi M_{\mathrm{S}}(T)$ on the assumption that $\Delta H_{q} = const$. (a) and (b) are determined for the 53\,$\mathrm{\mu m}$-thick sample. (c) Normalized relaxation rate obtained for YIG films of the thicknesses of 5.6\,$\mathrm{\mu m}$, 6.7\,$\mathrm{\mu m}$ and 53\,$\mathrm{\mu m}$ epitaxially grown on a GGG substrate of 500\,$\mathrm{\mu m}$ thickness. (d)~Comparison of the normalized relaxation rates of 53\,$\mathrm{\mu m}$-thick YIG on GGG, 30\,$\mathrm{\mu m}$-thick substrate-free YIG and an ultrapure bulk YIG sample measured at $H=H_\mathrm{c}$. The number in brackets corresponds to the material layer thickness in micrometers.}
\end{figure*}

It is known that the saturation magnetization $4\pi M_{\mathrm{s}}$ for bulk YIG crystals demonstrates a rather non-linear change with temperature \cite{Anderson1964}, which can be calculated using the two-sublattice model described in Ref.\,\cite{Hansen1974} as it is shown by the red solid line in Fig.\,\ref{fig3}\,(a). By-turn, the temperature behavior of the cubic anisotropy field can be approximated \cite{Danilov1989} as 
\begin{equation} 
H^\mathrm{c}_\mathrm{a} = H^\mathrm{c}_\mathrm{a} (0) +\alpha T^{\frac{3}{2}},
\label{H_a}
\end{equation}
with $H^\mathrm{c}_\mathrm{a} (0) = -147\,\mathrm{Oe}$ and $\alpha = 2.17 5\cdot 10^{-2}\,\mathrm{Oe}\,\mathrm{K^{-1.5}}$.

The slopes of the $4 \pi M_{\mathrm{s}}(T) - H^\mathrm{c}_\mathrm{a}(T) - H^\mathrm{u}_\mathrm{a}(T)$ curves determined for all films at room temperatures are in good agreement with previously reported results \cite{Obry2012, Langner2017}.
However, due to the unknown contribution of the uniaxial anisotropy \cite{Kaack1999}, which is caused by a temperature dependent mismatch between YIG and GGG crystal lattices, the calculated temperature dependencies for both $4\pi M_{\mathrm{s}}$ and for $4 \pi M_{\mathrm{s}} - H^\mathrm{c}_\mathrm{a}$ (dashed blue line in Fig.\,\ref{fig3}\,(a)) significantly diverge from the experimental difference $4 \pi M_{\mathrm{s}} - H^\mathrm{c}_\mathrm{a} - H^\mathrm{u}_\mathrm{a}$ (see, e.g., the data for the 53\,$\mathrm{\mu m}$-thick YIG film shown by red circles in Fig.\,\ref{fig3}\,(a)).
At the same time, the substrate-free YIG sample of 30~$\mu \mathrm{m}$ thickness prepared from the 53\,$\mathrm{\mu m}$-thick YIG film demonstrates good agreement between experimentally measured (empty blue circles) and theoretically calculated values of $4 \pi M_{\mathrm{s}} - H^\mathrm{c}_\mathrm{a}$. In the case of the thinner YIG films the experimental data for $4 \pi M_{\mathrm{s}} - H^\mathrm{c}_\mathrm{a} - H^\mathrm{u}_\mathrm{a}$ follow the general trend of the calculated saturation magnetization $4\pi M_{\mathrm{s}}$ (see Fig.\,\ref{fig3}\,(a)). This agreement evidences the applicability of the chosen model for our YIG films and allows us to use the theoretical magnetization values for the calculation of the temperature dependent parametric coupling strength.


By assuming initially the value of $\Delta H_{q}$ in Eq.\,\ref{h_thr} to be constant over the entire temperature range and taking into account the theoretical values of $4\pi M_{\mathrm{s}}(T)$ we have calculated the normalized (with respect to 340\,K) temperature dependence of the threshold field, which is solely determined by the change in the parametric coupling strength. This dependence is shown by the circles in Fig.~\ref{fig3}\,(b). 

The \textit{experimental} threshold field $h_{\mathrm{thr}}^\mathrm{exp}$, which contains information about the relaxation of parametrically excited magnons, can be found from the \textit{measured} threshold powers using the relation $h_\mathrm{thr}^\mathrm{exp}=C \sqrt{P_{\mathrm{thr}}}$. The value of $C$ depends on the pumping frequency $\omega_{\mathrm{p}}$, the geometry and the quality factor of the pumping resonator. As the resonance frequency and the quality factor of our microstrip resonator do almost not change with temperature we assume $C$ to be constant. 

The experimental values of the dimensionless threshold field normalized to the reference value at the temperature of 340\,K are plotted in Fig.~\ref{fig3}\,(b) (squares) for magnons excited near the FMR frequency ($H=H_\mathrm{c}$). Its behavior is visibly non-monotonic: down to 180\,K the threshold field $h_{\mathrm{thr}}^\mathrm{exp}$ slightly decreases, while below 180\,K it increases up to 6.5 times compared to the reference value. \looseness=-1

The comparison of the calculated ($h_{\mathrm{thr}}(T)$, circles) and experimental ($h_{\mathrm{thr}}^\mathrm{exp}(T)$, squares) threshold dependencies clearly evidences that at high temperatures  ($T \ge 180$\,K) the experimental dependencies are mostly determined by the variation in the parametric coupling strength. 
On the contrary, the strong increase of $h_{\mathrm{thr}}^\mathrm{exp}$ in the low-temperature range ($T<180$\,K) is caused by the spin-wave relaxation. 

Figure\,\ref{fig3}\,(c) shows the normalized relaxation rate $\Gamma_{q}$ calculated at $H_{\mathrm{c}}$ with help of Eq.\,\ref{h_thr} using $h_{\mathrm{thr}}^\mathrm{exp}(T)$ and theoretically calculated $4\pi M_{\mathrm{s}}(T)$. 
It becomes evident that for the temperature decrease from 180\,K to 20\,K the relaxation rate $\Gamma_{q}$ increases up to about 10.5 times for the 53\,$\mu$m-thick YIG film while the thinner films exhibit the same trend. The same relaxation behavior, as it is clear from the nearly wavenumber-independent vertical shift of all threshold curves (see, e.g., Fig.\,\ref{fig2}), is observed in all range of probed magnon wavenumbers up to $6\cdot 10^5\,\mathrm{rad\,cm}^{-1}$.
The strong increase of the relaxation rate is considered to be atypical for \textit{pure} YIG, for which a monotonic decrease of $\Gamma_{q}$ is expected with decreasing temperature \cite{Sparks}. 

The revealed relaxation behavior at low temperatures can be related either to the contribution of fast-relaxing rare-earth ions contaminating the chemical composition of YIG \cite{Dionne2000, Hansen1983} or to the magnetic losses caused by the dipolar coupling of magnons with the spin-system of the paramagnetic GGG substrate \cite{Balinskii1986,Danilov1996}.
In order to clarify the origin of the increased relaxation we replicate our measurements on the 53~$\mu \mathrm{m}$-thick YIG sample after polishing the GGG side down to a 30~$\mu \mathrm{m}$ substrate-free YIG film. The comparison of the relaxation rate is shown in Fig.\,{\ref{fig3}}\,(d).
Both YIG film samples demonstrate a strong increase in the magnon relaxation rate for decreasing temperatures, starting from approximately 150\,K. This fortifies the assumption of the prevailing contribution of fast-relaxing rare-earth ion impurities in epitaxial YIG films at low temperatures. 
Below approximately 80\,K the relaxation rate of the 53~$\mu \mathrm{m}$-thick YIG sample increases faster in comparison with the polished substrate-free YIG film. This difference can be attributed to coupling of YIG's ferrimagnetic spin system with the electron spins of $\mathrm{Gd^{3+}}$ ions of paramagnetic GGG. The coupling is supposed to be proportional to $1/T$ \cite{Balinskii1986} and leads to additional low-temperature energy losses for all YIG films placed on GGG substrates. 

For comparison, we measured the temperature-dependent magnon damping in an impurity- and GGG-free bulk YIG sample. Similar to the experiments with YIG films, these measurements, which were performed by means of the parallel parametric pumping technique in an ultrapure YIG crystal of the size of $(1 \times 1 \times 3)~\mathrm{mm}^3$, show the same damping behavior in a wide range of magnon wavenumbers. 
However, in contrast to the experiment with YIG films the relaxation rate  $\Gamma_{q}$ monotonically decreases with decreasing temperature. It is clearly shown by the line denoted by semi-filled circles in Fig.\,{\ref{fig3}}\,(d). 
It further supports our assumption about the significant influence of chemical contaminations on magnon damping in epitaxial YIG films at cryogenic temperatures.


In conclusion, in the temperature range from 20\,K to 340\,K we have investigated the relaxation of parametrically excited dipole-exchange magnons in YIG films of 5.6\,$\mathrm{\mu m}$, 6.7\,$\mathrm{\mu m}$, and 53\,$\mathrm{\mu m}$ thickness grown on a GGG substrate by liquid phase epitaxy. We have found that at cryogenic temperatures the magnon lifetime strongly decreases for all film thicknesses. By comparing the substrate-free YIG with the YIG/GGG samples the observed relaxation behavior could be related to the magnetic damping caused by the coupling of magnons to fast-relaxing rare-earth ions inside the YIG film ($T\lesssim 150$\,K) and to the paramagnetic spin system of GGG substrates ($T\lesssim 80$\,K). 
Comparison of these results with the data obtained from the ultrapure bulk YIG crystal shows that in order to sustain a long magnon lifetime low-temperature magnetic experiments in YIG must be performed in chemically-pure and substrate-free samples. 

Financial support from the Deutsche Forschungsgemeinschaft (project INST 161/544-3 within SFB/TR\,49, projects VA 735/1-2 and SE 1771/4-2 within SPP 1538 ``Spin Caloric Transport'', and project INST 248/178-1) is gratefully acknowledged.

\end{document}